\documentclass{article}  

\usepackage{spconf,amsmath,graphicx} 
\usepackage{enumitem}
\usepackage{amssymb}
\setlist{nosep, leftmargin=14pt}
\usepackage{mwe} 
\usepackage{caption} 
\usepackage[noadjust]{cite}  
\begin{document}

\title{Masked Autoencoders for Low dose CT denoising}
\name{Dayang Wang, Yongshun Xu, Shuo Han, Hengyong Yu*\thanks{*email: hengyong-yu@ieee.org. This work has been submitted to the IEEE for possible publication. Copyright may be transferred without notice, after which this version may no longer be accessible.}}


\address{Department of Electrical and Computer Engineering, \\ University of Massachusetts Lowell, Lowell, MA, USA, 01854.}

\maketitle

\begin{abstract}

Low-dose computed tomography (LDCT) reduces the X-ray radiation but compromises image quality with more noises and artifacts. A plethora of transformer models have been developed recently to improve LDCT image quality. However, the success of a transformer model relies on a large amount of paired noisy and clean data, which is often unavailable in clinical applications. In computer vision and natural language processing fields, masked autoencoders (MAE) have been proposed as an effective label-free self-pretraining method for transformers, due to its excellent feature representation ability. Here, we redesign the classical encoder-decoder learning model to match the denoising task and apply it to LDCT denoising problem. The MAE can leverage the unlabeled data and facilitate structural preservation for the LDCT denoising model when there are insufficient ground truth data. Experiments on the Mayo dataset validate that the MAE can boost the transformer's denoising performance and relieve the dependence on the ground truth data.
\end{abstract}
\section{Introduction}
In recent years, LDCT has become the mainstream in clinical applications due to the substantial x-ray radiation danger in normal-dose CT (NDCT). However, LDCT images compromise the image quality and diagnosis value, which has been a barrier for its applications. To overcome this issue, numerous deep learning models are explored along this direction \cite{chen2017low,RN72,shan2019competitive,zhang2021transct,wang2022ctformer}.

\textbf{Transformer Models.}
In the past few years, transformer models have gained lots of attention due to their ability to capture global contextual information. Several researches also cultivated their applications to the LDCT denoising. Zhang \textit{et al.} employed a Gaussian filter to decompose LDCT images into high/low frequency parts \cite{zhang2021transct}. Then, a transformer module was applied to the two parts for feature inference and contextual information fusion. Wang \textit{et al.} proposed a more advanced convolution-free encoder-decoder transformer. Then, the static and dynamic latent learning behavior of the model was revealed by analyzing the attention maps \cite{wang2022ctformer}. Recently, the Swin transformer has been very popular as a backbone architecture for a variety of downstream tasks \cite{RN74,wang2022sq}. The SwinIR is one of its important adaptions for image denoising, super-resolution, and artifact reduction \cite{liang2021swinir}. Notably, some SwinIR-based transformer models or modules have also been proposed for CT image enhancement/denoising and have achieved excellent results \cite{puttaguntaa2022swinir}. Therefore, the application of the MAE in this paper is also based on the SwinIR. 




\textbf{Masked Autoencoder.}
Recently, MAE has emerged as an excellent self-supervised learning strategy for various computer vision tasks \cite{he2022masked}. 
He \textit{et. al.} revealed that small fraction of an image can infer complex and holistic visual concepts like semantics \cite{he2022masked}. 
Zhou \textit{et. al.} showed that MAE can enhance medical image segmentation and classification performance \cite{zhou2022self}. However, these works are on high-level vision tasks. Currently, there is no exploration of MAE as self-pretraining on low-level vision tasks like LDCT denoising. 

Therefore, we are motivated to explore the potential of the MAE in LDCT denoising. We believe this study is significant for two reasons: i) There are usually insufficient ground truth data for clinical applications, and the self-pretraining paradigm of the MAE can reap the benefits of the unlabeled data. Therefore, it ought to be the ideal choice for an LDCT denoising work. ii) Structural preservation and enhancement are crucial goals in LDCT denoising. But, the anatomical structures in the CT image are supposed to be connected with each other mechanically and functionally. The MAE can aggregate the contextual information to infer the masked structures. Therefore, it can potentially strengthen the dependence between each anatomical region and complement the structural loss.


\section{Methods}

\subsection{Masked Autoencoder Design}
\begin{figure}
\centering
\includegraphics[width=\linewidth]{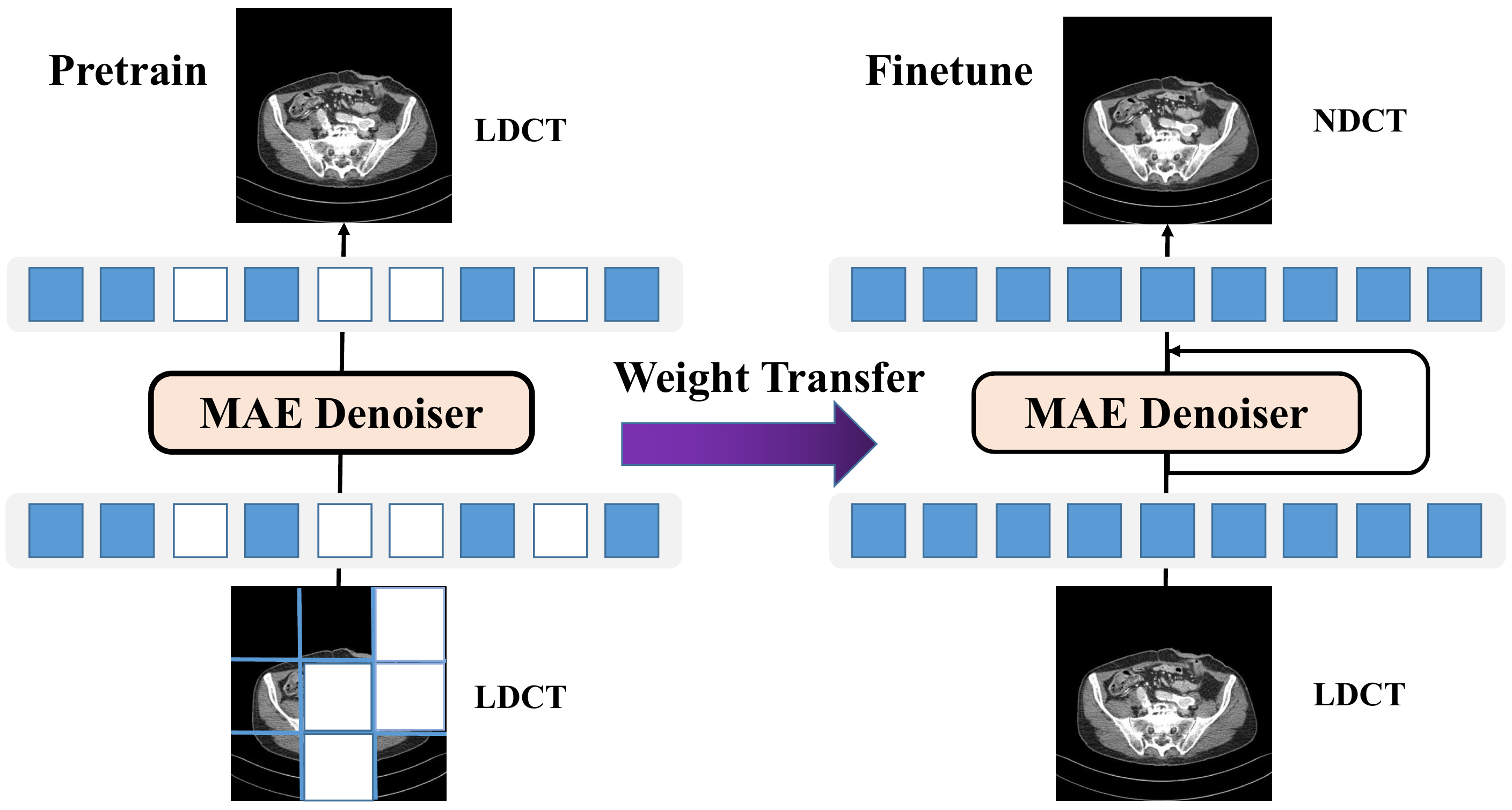}
\caption{The pipeline of the MAE in LDCT denoising.} \label{wholeflow1}
\end{figure}

As shown in the model flowchart in Fig. \ref{wholeflow1}, the MAE learning paradigm is employed for LDCT denoising. In the self-pretraining stage, the CT images are trained from LDCT images to LDCT images to learn the structure relationship, and the input is patch-wisely masked with a high rate (\textit{e.g.} $75\%$). Then, it is fed into the encoder and decoder to reconstruct the original image. In the authentic MAE design, the decoder is auxiliary and can be very lightweight to reduce the pretraining budget. Finally, in the finetuning process, the decoder is replaced with a task-specific module for a downstream task. 

Distinctively, we further ambiguate the encoder and decoder design by using a single MAE denoiser, because the LDCT denoising model shares the same image-to-image training paradigm as MAE pretraining. However, there is still a major difference between the two tasks since the MAE pretraining aims to complement the missing structures while the denoiser focuses on the Gaussian/Poisson noise removal. To accommodate for such discrepancies, we are inspired to redesign the pretraining model by adding two residual shortcuts and directly apply to the denoising task. In the finetuning stage, the model is trained from LDCT image to NDCT image to learn the noise removal process. Specifically, the image masks are removed and the model weights are transferred from the pretraining model with the identical design.

\subsection{Swin Transformer}
\begin{figure}
\centering
\includegraphics[width=\linewidth]{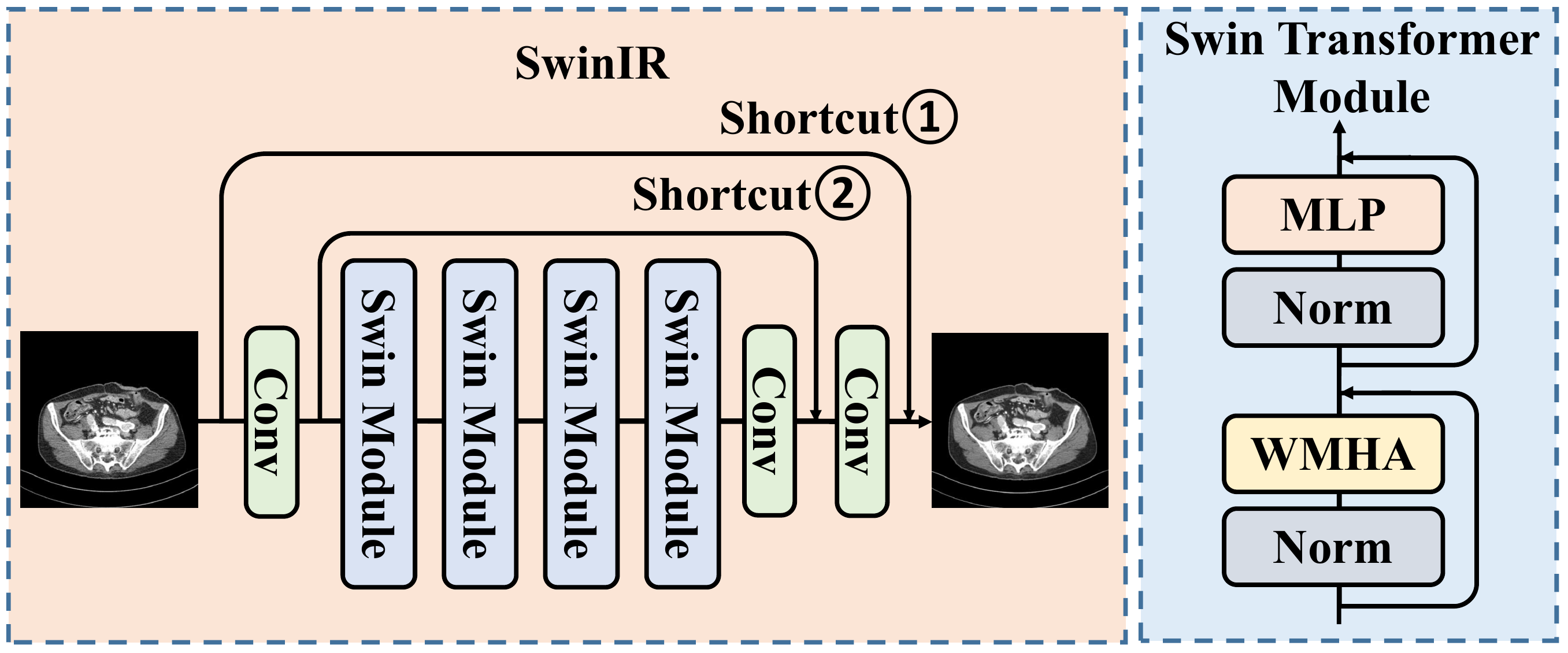}
\caption{The architecture of the SwinIR as MAE denoiser. 
Shortcuts \textcircled{\raisebox{-0.9pt}{1}} and \textcircled{\raisebox{-0.9pt}{2}} are removed in the MAE pretraining stage and then are reconnnected in the finetuning stage.}
\label{swin}
\end{figure}

Recently, The Swin transformer introduced hierarchical attention with shifted windows to facilitate contextual information fusion and reduce the high computational cost. SwinIR is featured as one of its most important applications for image denoising \cite{liang2021swinir}. As shown in Fig. \ref{swin}, it includes several convolutional blocks and Swin transformer modules. Specifically, for a Swin transformer module, an input token sequence $\mathbf{T_o} \in \mathbb{R}^{b \times w \times n \times d_o}$ is first layer-normalized (LN). Here $b$ is the batch size, $w$ is the window number, $n$ is the number of tokens, and $d_o$ is the token embedding dimension. Then, $\mathbf{T_o}$ is linearly transformed to query (Q), key (K), and value (V) to calculate the window-based multiple head attention (WMHA):
%
\begin{equation}
    \mathrm{WMHA}(\mathbf{Q},\mathbf{K},\mathbf{V}) = \mathrm{softmax}(\frac{\mathbf{Q}\mathbf{K^\top}}{\sqrt{d_k}}) \mathbf{V},
\end{equation}
where $\frac{1}{\sqrt{{{d}_{k}}}}$ is the scaling factor based on the network depth. Then, another LN layer, a multiple layer perceptron (MLP), and two residual shortcuts are applied to enrich feature inference. 
As shown in Fig. \ref{swin}, a Swin transformer module is characterized as
\begin{equation}
\begin{cases}
   & \mathbf{T}^{'}=  \mathrm{WMHA}(\mathrm{LN}( \mathbf{T_o})) + \mathbf{T_o} \\
   & \mathbf{T}_{\mathrm{s}}= \mathrm{MLP}(\mathrm{LN}(\mathbf{T}')) + \mathbf{T}^{'},
\end{cases}
\end{equation}
where $\mathbf{T'}$, $\mathbf{T_s}\in \mathbb{R}^{b \times w \times n \times d}$ are intermediate and final module outputs, respectively.

\subsection{Noise Model}
There are mainly two types of noises in the LDCT image. The first type is the thermal noise generated by the detection system, and this noise obeys the Gaussian distribution that is structurally independent. The second type of noise is photon statistical noise caused by unavoidable quantum fluctuations during the X-ray emission. This noise typically follows a Poisson distribution that is associated with the object structures. The LDCT noise follows a complex statistical distribution combining the Gaussian and Poisson noises: 
\begin{equation}
    \mathbf{p} \thicksim  \mathcal{N}(0,\sigma^2) +  \mathcal{P} (\xi(\boldsymbol{x})) ,
\end{equation}
where $\mathbf{p}$ is the noise variable, $\mathcal{N}(0,\sigma^2)$ denotes the Gaussian distribution whose variance is $\sigma^2$, and $\mathcal{P} (\xi(\boldsymbol{x}))$ is Poisson distribution in which  $\xi(\boldsymbol{x})$ denotes the latent mapping from image structure $\boldsymbol{x}$ to the distribution expectancy.

The proposed deep learning-based LDCT denoising model learns from a paired noisy LDCT image to a clean NDCT image directly. Particularly, $\mathbf{L}_1$ and $\mathbf{SSIM}$ losses are used to suppress these noises during model training. The $\mathbf{L}_1$ loss tends to remove the Gaussian noise more and preserves the brightness and color of the images. The $\mathbf{SSIM}$ loss will reduce the Poisson noise by keeping the contrast in the high-frequency area, thus enhancing the key structures.

\section{Experiments and Results}


\begin{figure}
\centering
\includegraphics[width=\linewidth]{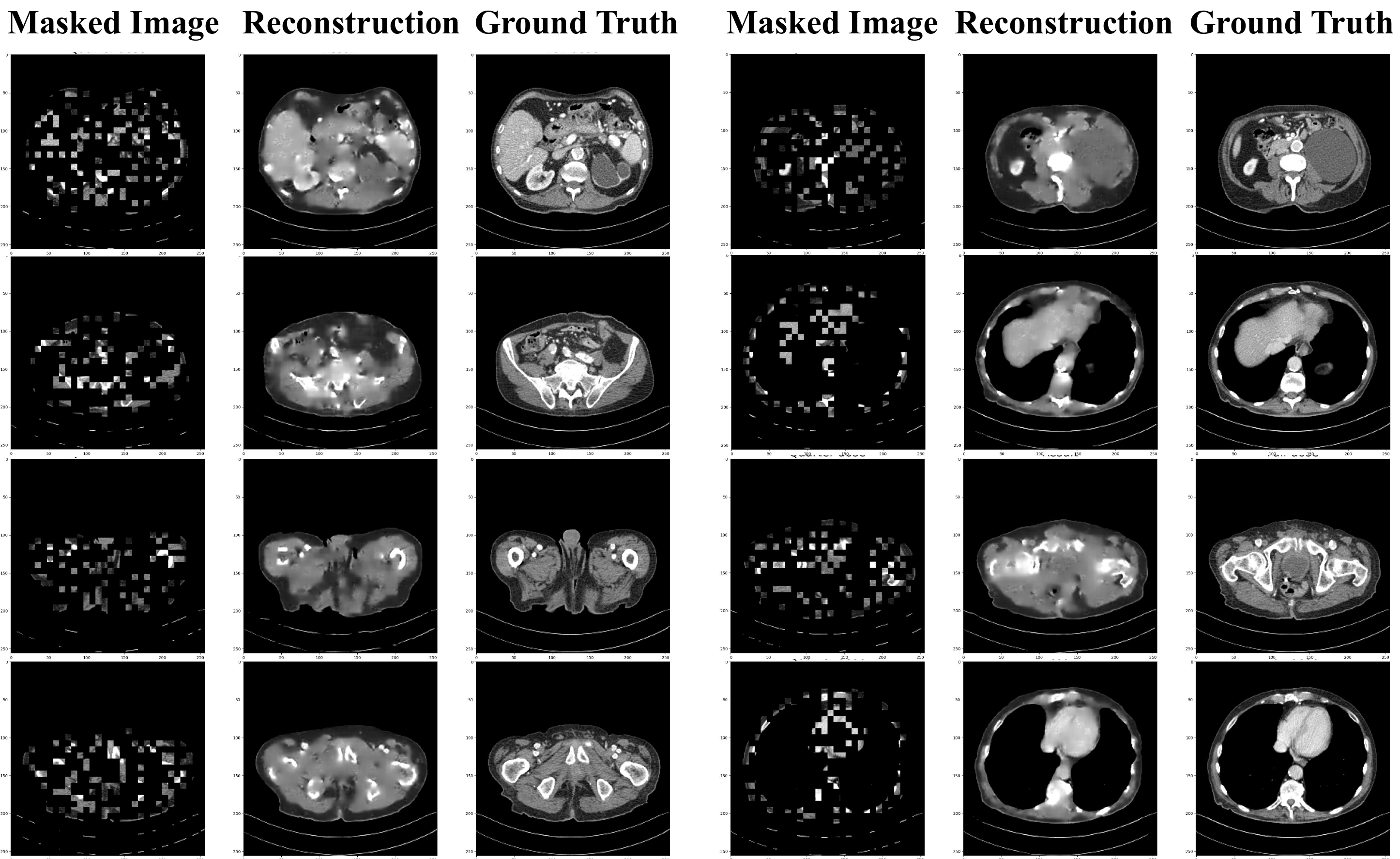}
\caption{The MAE reconstruction results on the testing cases.} 
\label{mae_result}
\end{figure}

\subsection{Dataset}
The publicly released dataset from \textit{2016 NIH-AAPM-Mayo Clinic LDCT Grand Challenge}\footnote{https://www.aapm.org/GrandChallenge/LowDoseCT/} \cite{RN83} is utilized for model training and evaluation. The dataset includes $2,378$ $3.0$mm slice thickness of CT images from ten anonymous patients. Ten-fold cross-validation is used for training and testing, with each fold taking one patient for testing and the other nine for training. We resized the original $512\times512$ images to $256\times256$ with \textit{inter\_area} interpolation. Then, the image pairs of quarter-dose LDCT and full-dose NDCT images are used to train the model.

\subsection{Experiment Settings}
The experiments are run on Ubuntu 18.04.5 LTS, with Intel(R) Core (TM) i9-9920X CPU @ 3.50GHz. All models are run with PyTorch 1.5.0 and CUDA 10.2.0 on one NVIDIA 2080TI 11G GPU. Below are the experiment settings: 

\begin{itemize}
    \item The masked patch size is 8 with a masked rate of $75\%$ during the self-pretraining process. 
    \item The model depths for the SwinIR are [4,4,4,4] with a uniform embedding dimension of 60 and 6 heads for the multiple head attention. 
    \item Both the MAE pretraining and the finetune processes are optimized through 50 epochs with the Adam optimizer. The initial learning rate is 1.5e-4 with a scheduled learning rate decay of 0.5 every 3000 iterations. 
    \item The batch size is 1 for both MAE pretraining and finetuning processes.  
\end{itemize}



\subsection{MAE Reconstruction Result}
In the self-pretraining process, the MAE learns a mapping from the corrupted image to the original image. Representative results on testing case slices are visualized in Fig. \ref{mae_result}. We can see that a small portion ($25\%$) of an original image can reconstruct the entire legitimate image to a great extent. Except that a few details are missing, the overall structures and anatomical regions can be retrieved after the pretraining.

\subsection{Comparison Results}
To validate the effectiveness of MAE pretraining for the transformer denoising model, the following state-of-the-art models are investigated: REDCNN \cite{chen2017low}, MAPNN \cite{shan2019competitive}, and SwinIR \cite{liang2021swinir}. Particularly, all these methods are implemented and optimized according to the officially disclosed codes. In addition, the MAE is performed as a pretraining strategy on the SwinIR (SwinIR+MAE). These four methods are trained on nine labeled patient data in a fully supervised way. Moreover, since the ground truth data is not always attainable in real-world clinical applications, we also explored SwinIR+MAE in a semi-supervised context, where three labeled patient data and six unlabeled patient data are used for training. 
\begin{figure}
\centering
\includegraphics[width=\linewidth]{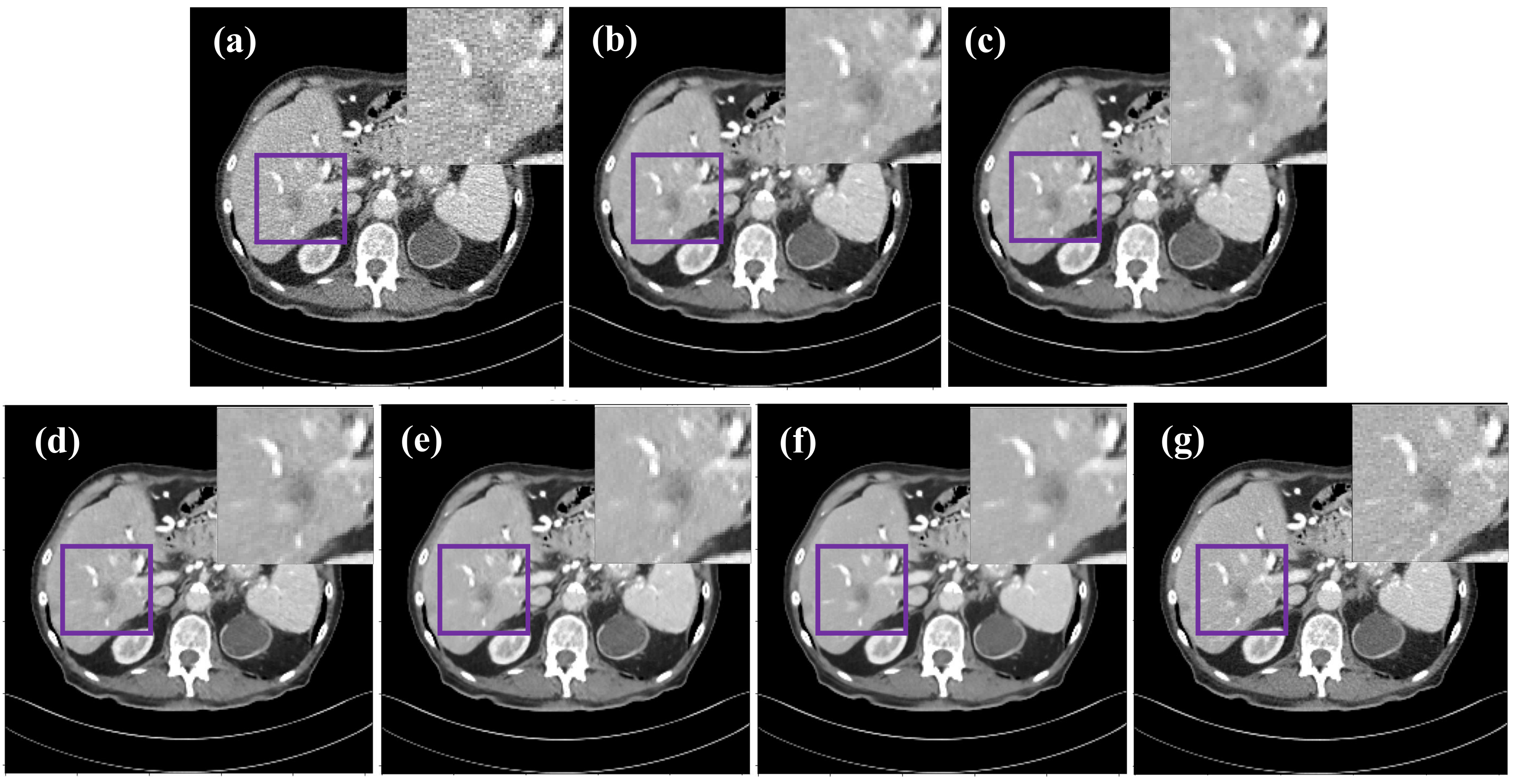}
\caption{The results of the model on Lesion 575 from patient L506. (a) LDCT, (b) REDCNN, (c) MAPNN, (d) SwinIR, (e) SwinIR+MAE(semi), (f) SwinIR+MAE, (g) NDCT. The display window is [-160,240] \textit{HU}.} 
\label{86whole}
\end{figure}
Results from Fig. \ref{86whole} show that all these methods are effective in removing the noises from LDCT images. The proposed SwinIR+MAE can generate clearer and smoother images without losing any key structures compared to other models. Furthermore, compared to supervise-trained SwinIR, SwinIR+MAE(semi) design can produce CT images with less blotchy artifacts and higher image quality without any structural distortion. 

Furthermore, our model is also assessed quantitatively with two metrics including \textit{structural similarity} (SSIM) and \textit{root mean square error} (RMSE). Results in Tab. \ref{tab:quant} indicate that the SwinIR models can achieve better performance than REDCNN and MAPNN. Next, the introduction of the MAE to the SwinIR in supervised design delivers the best scores with SSIM of 0.9609 and RMSE of 6.7355. Moreover, it can also achieve high performance in the semi-supervised learning context, even higher than the raw SwinIR in full-supervised training.

\begin{table}[h]
\caption{Quantitative results of different methods on L$506$. The bold-faced numbers are the best results. }
    \centering
    \scalebox{.92}{
    \begin{tabular}{|l|c|c|}
    \hline
    Method  &  SSIM$\uparrow$  &   RMSE$\downarrow$   \\
    \hline
    LDCT      &0.9359  & 10.4833 \\
    REDCNN     &0.9501$\pm{0.0012}$  & 7.8016$\pm{0.0834}$ \\
    MAPNN      &0.9504$\pm{0.0019}$  & 7.5357$\pm{0.1342}$ \\
    SwinIR   & 0.9546$\pm{0.0011}$ & 7.2232$\pm{0.1084}$\\
    SwinIR+MAE(semi)   & 0.9601$\pm{0.0009}$ & 6.8326$\pm{0.0780}$\\
    SwinIR+MAE  & \textbf{0.9609$\pm{0.0007}$} & \textbf{6.7355$\pm{0.0800}$}\\
    \hline
    \end{tabular}}
  \label{tab:quant}
    \hfill
\end{table}

\section{Conclusion}
In this paper, the application of masked autoencoder in the LDCT is studied. To the best of our knowledge, this is the first time to explore the MAE as a self-pretraining strategy in the low-level denoising task. Experimental results on the AAPM dataset showed that the MAE can enhance the model in noise removal and structural preservation. It can generate more perceptually pleasing CT images qualitatively and quantitatively. Moreover, it can also facilitate a semi-supervised learning context with higher scores than the fully supervised scenario, thus relieving the ground truth data dependence in real applications. In the future, we will further explore the application of the MAE to other medical image modalities.

\section{Compliance with Ethical Standards}
This research study was conducted retrospectively using human subject data made available in open access by Mayo clinic \cite{RN83}. Ethical approval was not required as confirmed by the license attached with the open access data.

\section{Acknowledgments}
No funding was received to conduct this study. The authors have no relevant financial/non-financial interests to disclose.
\bibliographystyle{IEEEbib}
\bibliography{ISBIMAE}

\end{document}